\documentclass[twocolumn]{aastex631}

\usepackage{multirow}
\usepackage{booktabs}
\usepackage{threeparttable}
\usepackage{ulem}

\shorttitle{GRB 170519A}
\shortauthors{Z. M. Zhou et al.}
\graphicspath{{./}{figures/}}
\begin{document}
\title{GRB 170519A: Thermal Radiation in an X-ray Flare and Decaying Magnetic Fields for the Early-Time Afterglow}


\author[0000-0003-3360-2211]{Zi-Min Zhou}
\affiliation{Guangxi Key Laboratory for Relativistic Astrophysics, School of Physical Science and Technology, Guangxi University, Nanning 530004, China; wangxg@gxu.edu.cn}

\author[0009-0007-7188-3196]{Liang-Jun Chen}
\affiliation{Guangxi Key Laboratory for Relativistic Astrophysics, School of Physical Science and Technology, Guangxi University, Nanning 530004, China; wangxg@gxu.edu.cn}

\author{Rui-Quan Li}
\affiliation{Guangxi Key Laboratory for Relativistic Astrophysics, School of Physical Science and Technology, Guangxi University, Nanning 530004, China; wangxg@gxu.edu.cn}

\author[0000-0001-8411-8011]{Xiang-Gao Wang}
\affiliation{Guangxi Key Laboratory for Relativistic Astrophysics, School of Physical Science and Technology, Guangxi University, Nanning 530004, China; wangxg@gxu.edu.cn}
\affiliation{GXU-NAOC Center for Astrophysics and Space Sciences, Nanning 530004, China}

\author{Xing-Ling, Li}
\affiliation{Guangxi Key Laboratory for Relativistic Astrophysics, School of Physical Science and Technology, Guangxi University, Nanning 530004, China; wangxg@gxu.edu.cn}

\author{En-Wei Liang}
\affiliation{Guangxi Key Laboratory for Relativistic Astrophysics, School of Physical Science and Technology, Guangxi University, Nanning 530004, China; wangxg@gxu.edu.cn}
\affiliation{GXU-NAOC Center for Astrophysics and Space Sciences, Nanning 530004, China}

\author[0000-0002-2636-6508]{WeiKang Zheng}
\affiliation{Department of Astronomy, University of California, Berkeley, CA 94720-3411, USA; weikang@berkeley.edu, afilippenko@berkeley.edu}

\author{Alexei V. Filippenko}
\affiliation{Department of Astronomy, University of California, Berkeley, CA 94720-3411, USA; weikang@berkeley.edu, afilippenko@berkeley.edu}

\begin{abstract}
GRB 170519A was discovered by \emph{Swift}/BAT, and then observed by \emph{Swift}/XRT, \emph{Swift}/UVOT, and ground-based telescopes. We report Lick/KAIT observations of GRB 170519A, and make temporal analysis and spectral joint fits of its multiwavelength light curves. The observations present a relatively complete afterglow structure, including two X-ray flares (Flares I and II), optical onset (Slice 1), normal decay (Slices 2 and 3), and a possible jet break. The spectrum of the bright X-ray flare (Flare II) indicates that a thermal component exists at $t = 190$--240~s. \textbf{The blackbody emits in the photospheric radius $R_{\rm ph}\sim 10^{11}$ cm,} and its temperature ($kT$) decreases with time from \textbf{1.08 to 0.37 keV, its Lorentz factor of blackbody ($\Gamma_{\rm BB}$) decreases with time from 67.71 to 46.70. The luminosity of the blackbody ($L_{\rm BB}$), $kT$ and $\Gamma_{\rm BB}$ follow the relations $\bf L_{\rm BB} \propto kT^{2.49\pm 0.03}$ and $\Gamma_{\rm BB}\propto L_{\rm BB}^{0.27}$ (estimated from \cite{fan2012}).} In the optical light curves, there is an onset bump in the early-time afterglow, rising with an index $\alpha_{O,1} \approx -0.43$ and peaking $\sim1174.9$ s since the BAT trigger. The bump then decays with $\bf \alpha_{O,2} \approx 0.88$ in the normal decay phase, and the X-ray flux decays with a similar index of $\bf \alpha_{X,1} \approx 0.95$. There is no obvious spectral evolution in the normal decay phases, with photon index $\hat{\Gamma} = 1.86$ and 1.92 in Slices 2 and 3, respectively. We find that the multiwavelength light curves of the GRB 170519A afterglow can be well fitted by an external shock with time-dependent $\epsilon_B$. In the early afterglow, the value of $\epsilon_B$ decays rapidly from $\bf 4.29\times10^{-2}$ to $\bf 8.23\times10^{-3}$.
\end{abstract}
\keywords{gamma-ray bursts: general --- gamma-ray bursts: individual (GRB 170519A)}

\section{Introduction} \label{sec:intro}
A gamma-ray burst (GRB) is one of the most luminous phenomena in the Universe \citep{piran1999, kumarzhang15, zhangbook18}, and is considered to originate from a relativistic jet powered by collapse of a massive star \citep{narayan1992, woosley1993, macfadyen1999, fruchter2006, zhang2006} or the merger of compact binaries (e.g., binary neutron stars, or neutron-star--black-hole system) \citep{paczynski1986, eichler1989, paczynski1991, zhang2006}. The most popular model for GRBs is the standard fireball model. During the expansion of the fireball, the internal shock is generated through collisions of relativistic shells, and it accelerates electrons in shell to produce nonthermal prompt $\gamma$-ray emission \citep{blinnikov1984, rees1994, kobayashi1997, daigne1998, rees1998, piran1999}. On the other hand, owing to the high electron density in a fireball, photons are repeatedly scattered and trapped until the optical depth decreases \citep[e.g.,][]{goodman1986, merzaros2000}. Hence, the thermal radiation may arise in the optically thin region \citep[e.g.,][]{ryde2010, axelsson2012, zhang2012, larsson2015, arimoto2016, hou2018}, and possibly hide in the intense nonthermal emission \citep{ryde2005, ryde2009, guiriec2011, axelsson2012, zhangbb2011, guiriec2013, lv2017}. 
As the jet moves farther away from the center and sweeps through the external medium, the shell interacts with the medium and produces an external shock which generates a multiwavelength afterglow \citep{merszaros1997, sari1998, gao2013}. 

Based on the features of the X-ray emission, the X-ray afterglow is commonly classified into five temporal components \citep{zhang2006, zhangbook18}: (1) steep decay phase, as the tail of the prompt emission \citep{barthelmy2005, tagliaferri2005, zhang2007}; (2) shallow decay phase \citep{Nousek2006, liang2007, wang2015}; (3) normal decay phase, which may be consistent with the external shock radiation from the fireball \citep[e.g.,][]{sari1998} (4) jet break in the late-time afterglow \citep[e.g.,][]{liang2008, racusin2009}; and (5) X-ray flare, which is often believed to have an internal origin and may be the signal of the restart of the central engine after the prompt gamma-ray phase \citep{burrows2005, ioka2005, fanwei2005, Liang2006, Chincarini2007, lazzati2007, maxham2009, margutti2010, Margutti2011, peng2014, lv2022}. The X-ray flare associated with GRBs is usually observed in the soft X-rays, and $\sim 33$\% of GRB afterglows exhibit flares \citep{Chincarini2010}. Extensive studies have been conducted on the spectral and temporal properties of X-ray flares \citep[e.g.,][]{OBrien2006, Liang2006, Chincarini2007, Falcone2007, Chincarini2010, Margutti2011, Qin2013, Wang2013, Hu2014, Yi2016}. The spectra of most flares can be adequately fitted with an absorbed single power-law (PL) model, although the GRB Band function \citep{Band1993} or the PL plus the blackbody (BB) radiation component may improve the fit for some flares \citep{Falcone2007, Page2011}. The simplest external forward-shock models can account for the multiwavelength afterglow during the shallow decay and normal decay phases \citep{wang2015}. In this model, the microphysical parameters (e.g., fractions of shock energy that go to electrons $\epsilon_e$ and magnetic fields $\epsilon_B$) are typically assumed to be constants during the late-time afterglow, though some investigations consider variations of the microphysical parameters during the early-time afterglow \citep{Yost2003, Ioka2006, Fan2006, Granot2006,Kong2010,Huang2018,Fraija2024}.

GRB 170519A was triggered by the \emph{Swift}/Burst Alert Telescope (BAT), and subsequently observed by the \emph{Swift}/X-Ray Telescope (XRT) as well as several optical telescopes. The event exhibits prompt emission, a multwavelength bright flare, and an afterglow (including onset, normal decay), providing sufficient data to adequately investigate the physics of GRBs. In this paper, we present our observations using the Katzman Automatic Imaging Telescope (KAIT), and combine data from other satellites to study the multiwavelength afterglow of GRB 170519A. Our observations are detailed in Section \ref{sec:obs}. In Section \ref{sec:analyses}, we perform the temporal analysis (Section \ref{sec:temporal}) and spectral analysis (Section \ref{sec:spectral}) for GRB 170519A. Based on the fitting of light curves and spectra, we further discuss the presence of a thermal component in its bright X-ray flare in Section \ref{sec:flare2}, and model the light curves with the external forward shock model in Section \ref{sec:modeling}. The results are summarized in Section \ref{sec:conclusions}. For the calculations of energy and luminosity, we use the luminosity distance $D_{\rm L}(z)$ in which the cosmological parameters $\Omega_{\rm{m},0} = 0.3$, $\Omega_{\Lambda,0} = 0.7$, and H$_{0} = 70$ km s$^{-1}$ Mpc$^{-1}$ are adopted.

\section{Observations}
\label{sec:obs}
\begin{figure}[htb!]
\centering
\includegraphics[angle=0,scale=0.3]{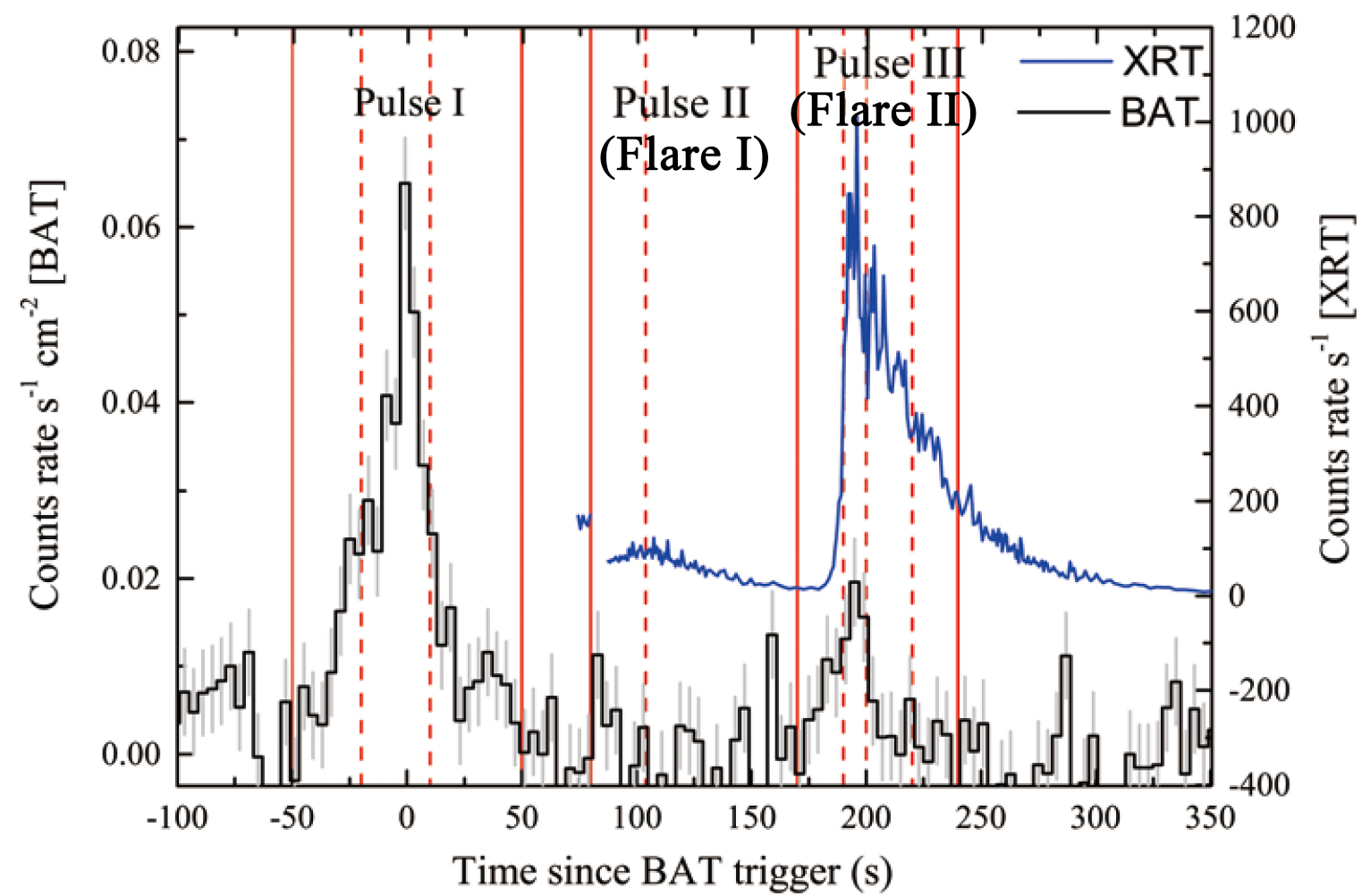}
\caption{Light curves from \emph{Swift}/BAT and \emph{Swift}/XRT.}
\label{fig:BAT-XRT-counts}
\end{figure}

GRB 170519A was triggered by the \emph{Swift}/BAT on 2017 May 19 at 05:10:02 UTC (denoted as $T_0$ in this paper), with $T_{90}=216.4\pm49.4$~s \citep{Krimm2017}. The XRT and the UltraViolet-Optical Telescope (UVOT) onboard \emph{Swift} started observing the X-ray and optical afterglows of GRB 170519A at 80.4 and 91.0~s since $T_0$, respectively. We downloaded the BAT and XRT data from the NASA \emph{Swift} archive\footnote{The light curves and spectra of XRT data are downloaded from \url{https://www.swift.ac.uk/xrt_curves/00753445/} and \url{https://www.swift.ac.uk/xrt_spectra/addspec.php?targ=00753445&origin=GRB}, respectively. \textbf{It is noted that, 4 XRT data points (in $T_0$ + [$6.72\times10^4$, $7.38\times10^4$] s, [$7.92\times10^4$, $9.69\times10^4$] s, [$1.60\times10^5$, $1.94\times10^5$] s, [$2.53\times10^5$, $4.65\times10^5$] s) may be unreliable (as the warning appeared on the website), so we did not include them in the X-ray light curve.} We obtained the BAT data from \url{https://www.swift.ac.uk/archive/selectseq.php?tid=00753445&source=obs}, and extract its light curves and spectra using the HEAsoft package \citep{Evans2009}.}. The light curves from \emph{Swift}/BAT and \emph{Swift}/XRT are shown in Figure \ref{fig:BAT-XRT-counts}. They exhibit three pulses: an obvious pulse in the prompt $\gamma$-ray emission detected by BAT (\textbf{Pulse I}), during $T_0$ + [$-50.0$, 50.0]~s; a soft X-ray \textbf{pulse} during the BAT quiescent period, $T_0$ + [80.0, 170.0]~s (\textbf{Pulse II}); and a pulse \textbf{detected with both XRT and BAT} during $T_0$ + [170.0, 240.0]~s (\textbf{Pulse III}).

GRB 170519A was rapidly monitored by several ground-based telescopes, including (1) iTelescope \citep{Izzo2017GCN.21108....1I,Hentunen2017GCN.21113....1H}, (2) the 1~m telescope located at Weihai, Shandong, China \citep{Xu2017GCN.21111....1X}, (3) the 2.5~m Nordic Optical Telescope at La Palma (Spain) \citep{deUgartePostigo2017GCN.21120....1D}, (4) the MITSuME 50~cm telescope of the Okayama Astrophysical Observatory \citep{Kuroda2017GCN.21121....1K}, (5) the 40~cm UCD Watcher telescope at Boyden Observatory in South Africa \citep{Martin-Carrillo2017GCN.21123....1M}, (6) the MITSuME 50~cm telescope of Akeno Observatory \citep{Morita2017GCN.21124....1M}, (7) the 22~cm SSS-220 telescope of ISON/Multa observatory \citep{Mazaeva2017GCN.21162....1M}, (8) the AZT-11 telescope of CrAO Observatory \citep{Mazaeva2017GCN.21169....1M}, (9) the AZT-33IK telescope of Sayan observatory \citep{Mazaeva2017GCN.21206....1M}, (10) the Zeiss-1000 1~m telescope of Tien Shan Astronomical Observatory \citep{Mazaeva2017GCN.21208....1M}, (11) the AMI Large Array \citep{Mooley2017GCN.21211....1M}, (12) OSIRIS on the 10.4~m GTC at the Roque de los Muchachos Observatory (which provided the redshift determination of $z = 0.818$; \citealt{Izzo2017GCN.21119....1I}), (13) the Reionization and Transients Infrared Camera \citep{Butler2017GCN.21109....1B, Butler2017GCN.21122....1B}, and (14) the 60~cm BOOTES-5/Javier Gorosabel Telescope at Observatorio Astron\'o micoNacional in San Pedro M\'artir (M\'exico) \citep{Castro-Tirado2017GCN.21117....1C}. Our own optical follow-up campaign of GRB 170519A was carried out using the 0.76~m Katzman Automatic Imaging Telescope \citep{filippenko2001} at Lick Observatory, starting at 05:14:24 UTC (262~s after the \emph{Swift} trigger) \citep{Zheng2017GCN.21115....1Z}. 

\begin{figure*}[htb!]
\centering
\includegraphics[angle=0,scale=0.5]{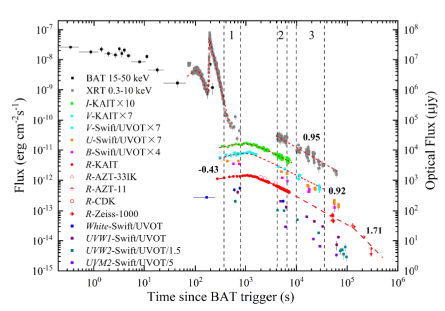}
\caption{Multiwavelength light curves of GRB 170519A on a logarithmic timescale.}
\label{LC-BAT-XRT-OPT}   
\end{figure*}

KAIT observations of GRB 170519A were performed with an automatic sequence in the $V$, $I$, and $Clear$ \citep[approximately equivalent to $R$, see][]{LiWeidong2003} filters, with an exposure time of 20~s per image, lasting approximately 2~hr. We also follow the UVOT analysis threads\footnote{\url{https://www.swift.ac.uk/analysis/uvot/index.php}} to create a light curve observed by \emph{Swift}/UVOT. The photometry obtained from KAIT, UVOT, and other optical telescopes collected through the General Coordinates Network are listed in Table \ref{tab:Photometry}. The multiwavelength light curve on a logarithmic timescale is shown in Figure \ref{LC-BAT-XRT-OPT}.

\startlongtable
\begin{deluxetable*}{cccccc}
\tabletypesize{\small}
\tablecaption{Optical Afterglow Photometry Log of GRB 170519A} \label{tab:Photometry}
\tablewidth{0pt}
\tablehead{\colhead{$T-T_{\rm 0}$ (s)$^{\rm a}$} & \colhead{Exp. (s)} & \colhead{Mag$^{\rm b}$} & \colhead{Mag $1\sigma$} & \colhead{Filter} & \colhead{Telescope$^{\rm c}$}}
\startdata
272	&	20	&	16.13	&	0.03	&	$Clear$	&	KAIT	\\
370	&	20	&	16.06	&	0.04	&	$Clear$	&	KAIT	\\
470	&	20	&	15.98	&	0.02	&	$Clear$	&	KAIT	\\
570	&	20	&	15.93	&	0.03	&	$Clear$	&	KAIT	\\
670	&	20	&	15.93	&	0.03	&	$Clear$	&	KAIT	\\
769	&	20	&	15.94	&	0.04	&	$Clear$	&	KAIT	\\
869	&	20	&	15.88	&	0.03	&	$Clear$	&	KAIT	\\
969	&	20	&	15.87	&	0.02	&	$Clear$	&	KAIT	\\
1069	&	20	&	15.84	&	0.03	&	$Clear$	&	KAIT	\\
...	&	...	&	...	&	...	&	...	&	...	\\
\enddata
\tablenotetext{a}{$T-T_{\rm 0}$ is the exposure median time after the BAT trigger.}
\tablenotetext{b}{Magnitudes are not corrected for Galactic extinction.}
\tablenotetext{c}{References: (1) \citet{2017GCN.21113....1H},(2) \citet{Mazaeva2017GCN.21169....1M}, (3) \citet{Mazaeva2017GCN.21208....1M}, (4) \citet{Mazaeva2017GCN.21206....1M}.}
\tablenotetext{}{(The full table is available in the electronic version.)}
\label{magnitude}
\end{deluxetable*}

\section{Temporal and Spectral Analyses}\label{sec:analyses}
\subsection{Behavior of Multiwavelength Afterglow}\label{sec:temporal}
To obtain the temporal profile of the afterglow of GRB 170519A, \textbf{we fit the light curves by employing the single power-law (SPL)} function \citep[e.g.,][]{liang2007, li2012, wang2015},
\begin{equation}
\bf
F_1 = F_{01} t^{-\alpha_1}, \label{SPL_LC}
\end{equation}
or the broken power-law (BPL) function,
\begin{equation}
F_2 = F_{02} \left[\left(\frac{t}{t_{\rm b}}\right)^{\alpha_1\omega} +\left(\frac{t}{t_{\rm b}}\right)^{\alpha_2\omega}\right]^{-1/\omega}, \label{BPL}
\end{equation}
where $\alpha_1$ and $\alpha_2$ are the temporal slope indices, $t_{\rm b}$ is the break time, and $\omega$ represents the sharpness of the break. 
The best fitting results for the SPL or BPL functions are presented in Table \ref{tab:bpl_tpl} and Figure \ref{LC-BAT-XRT-OPT}.

For the X-ray light curve, \textbf{Pulse II and Pulse III} can be fitted with a BPL function, with break times at $T_0$ + 114.8~s and $T_0$ + 190.6~s, respectively. \textbf{Pulse II exhibits rapid rise ($\alpha_1$= -1.25) and rapid steep decay ($\alpha_2$=4.40). Pulse III also exhibits rapid rise ($\alpha_1$= -100.00) and rapid decay ($\alpha_2$=7.24). These characteristics of the two pulses are consistent with typical X-ray flare \citep[e.g.,][]{burrows2005, Falcone2007, li2012,wang2015}. To investigate the their physical origin, we named the Pulse II and Pulse III as Flare I and Flare II, respectively. In the late phase (approxmately $T_0$ + [$4.00\times 10^2$, $1.00\times10^5$] s)}, the X-ray flux decreases with a power-law index $\bf \alpha_{\rm X,1}= 0.95$. 
For the optical afterglow \textbf{in $T_0$ + [20.0, $\bf 1.26 \times 10^5$] s}, a smooth onset bump rises with an index $\alpha_{\rm O,1} = -0.43$ and peaks at $\sim T_0$ + 1174.9~s, and then decreases with a power-law index of $\bf \alpha_{\rm O,2} = 0.92$, \textbf{finally decay from $\sim T_0$ + $1.26 \times 10^5$~s with a stepper index $\bf \alpha_{\rm O,3} \approx 1.71$ (fixed), which may correspond to the post-jet-break phase.} $\bf \Delta \alpha = \alpha_{\rm X,1}-\alpha_{\rm O,2} \approx 0$ indicates that the X-ray and optical data are located in the same spectral regime. 
 To further illustrate this point, we discuss the closure relation (the relation between power-law index of the light curve and spectrum) in Section \ref{sec:spectral}.

\startlongtable
\begin{deluxetable*}{ccccc}
\centering
\tablecaption{Temporal Parameters of GRB 170519A.} \label{tab:bpl_tpl}
\tablehead{\colhead{} &\colhead{$\alpha_1$} & \colhead{$\alpha_2$} & \colhead{$\alpha_3$} &\colhead{log $t_{\rm b}$ (s)} }
\startdata
Pulse II & $-1.25\pm0.42$ & $4.40\pm0.28$ & ... & $2.06\pm0.01$  \\
Pulse III & $-100.00\pm7$ & $7.24\pm0.08$ & ... & $2.28\pm0.01$  \\
\textbf{Later X-ray} & $\bf 0.95\pm0.05$ & ... & ...& ...  \\
$I$ band & $-0.37\pm0.07$ & $0.84\pm0.03$ & ...& $3.06\pm0.03$  \\
$V$ band & $-0.43\pm0.05$ & $0.84\pm0.07$ & ... & $3.04\pm0.06$ \\
$R$ \textbf{band} & $\bf -0.43\pm0.08$ & $\bf 0.92\pm0.01$ & $\bf 1.72$ (fixed) & $\bf 3.07\pm0.03$ \\
\enddata
\end{deluxetable*}

\subsection{Spectral Analysis}
\label{sec:spectral}
To get more information, we analyze the spectra with each time interval denoted by dashed lines in Figures \ref{fig:BAT-XRT-counts} and \ref{LC-BAT-XRT-OPT}, and fit the spectra using Xspec of the HEAsoft tool.
The XRT data are fitted in the energy range of 0.6–10 keV\footnote{\bf As noted in \cite{valan2021} and \cite{valan2023}, the XRT data of GRB 170519A at $< 0.6$ keV has some calibration issue, the detail is described in \url{https://www.swift.ac.uk/analysis/xrt/digest_cal.php}.}, and the BAT data are fitted in the range of 15-150 keV. For X-ray and optical spectra, the photoelectric absorption of hydrogen in our Galaxy and the host galaxy \citep{Wilms2000} should be considered. The Galactic hydrogen column density is set as $2.37 \times 10^{20}$ cm$^{-2}$ in the direction of this burst \citep{Willingale2013}. \textbf{We obtained the hydrogen column density of the host galaxy ($N_H$) by fitting the data using \emph{ztbabs} model} \citep{Wilms2000} in Xspec. \textbf{The fitting result of $N_H$ is $(4.1 \pm 0.7)\times 10^{21}$ cm$^{-2}$ in $T_0$ + [80.0, 240.0] s (XRT-WT data was used), and is $(3.8 \pm 3.5)\times 10^{21}$ cm$^{-2}$ in $T_0$ + [$3.70 \times 10^2$, $3.60 \times 10^4$] s (XRT-PC data was used). The values of $N_H$ are similar to the results of automatic XRT averaged-spectra published online\footnote{\rm $N_H = (5.3 \pm 0.3) \times 10^{21}$ cm$^{-2}$ in $T_0$ + [86.0, 373.0] s (WT data) and $N_H = (3.9 \pm 1.0) \times 10^{21}$ cm$^{-2}$ in $T_0$ + [$3.79 \times 10^2$, $3.36\times 10^4$] s (PC data), which were published on \url{https://www.swift.ac.uk/xrt_spectra/00753445/}.}. }
%


For the optical data, we considered the dust extinction of the host galaxy. The extinction caused by dust grains in the host galaxy is characterized by the extinction curves of the Small Magellanic Cloud (SMC), whose standard value for $R_V$ (the ratio of the total to selective extinction) is $R_{V,{\rm SMC}} = 2.93$ \citep{PeiYiChuan1992}; we estimate the color index $E_{B-V} = 0.021 \pm 0.014$ mag for the host galaxy. 

\begin{figure}[htb!]
\centering
\includegraphics[angle=0,scale=0.35]{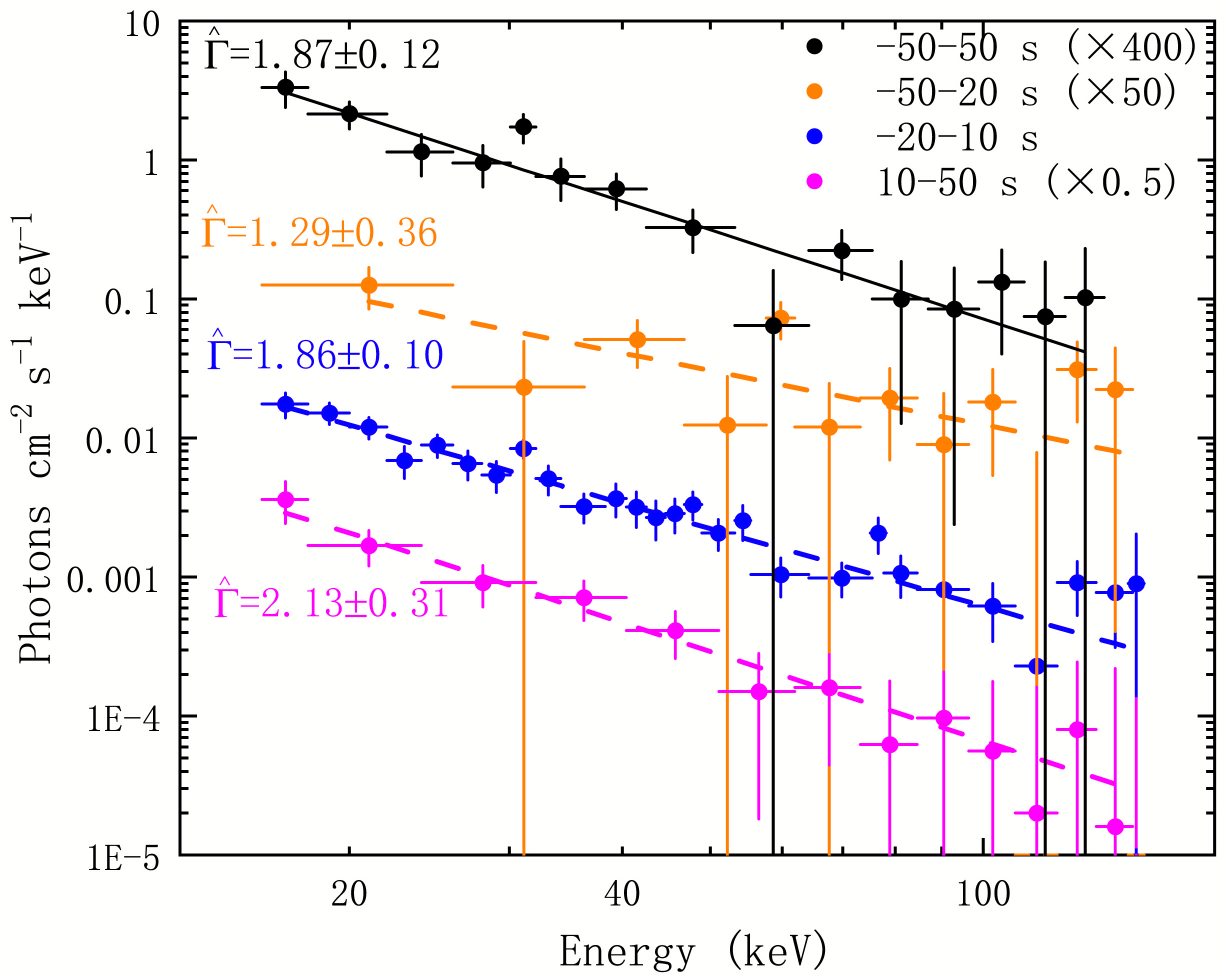}
\caption{Spectra of the first BAT pulse of GRB 170519A.}
\label{fig:BAT-spectra}
\end{figure}

\begin{figure}[htb!]
\centering
\includegraphics[angle=0,scale=0.35]{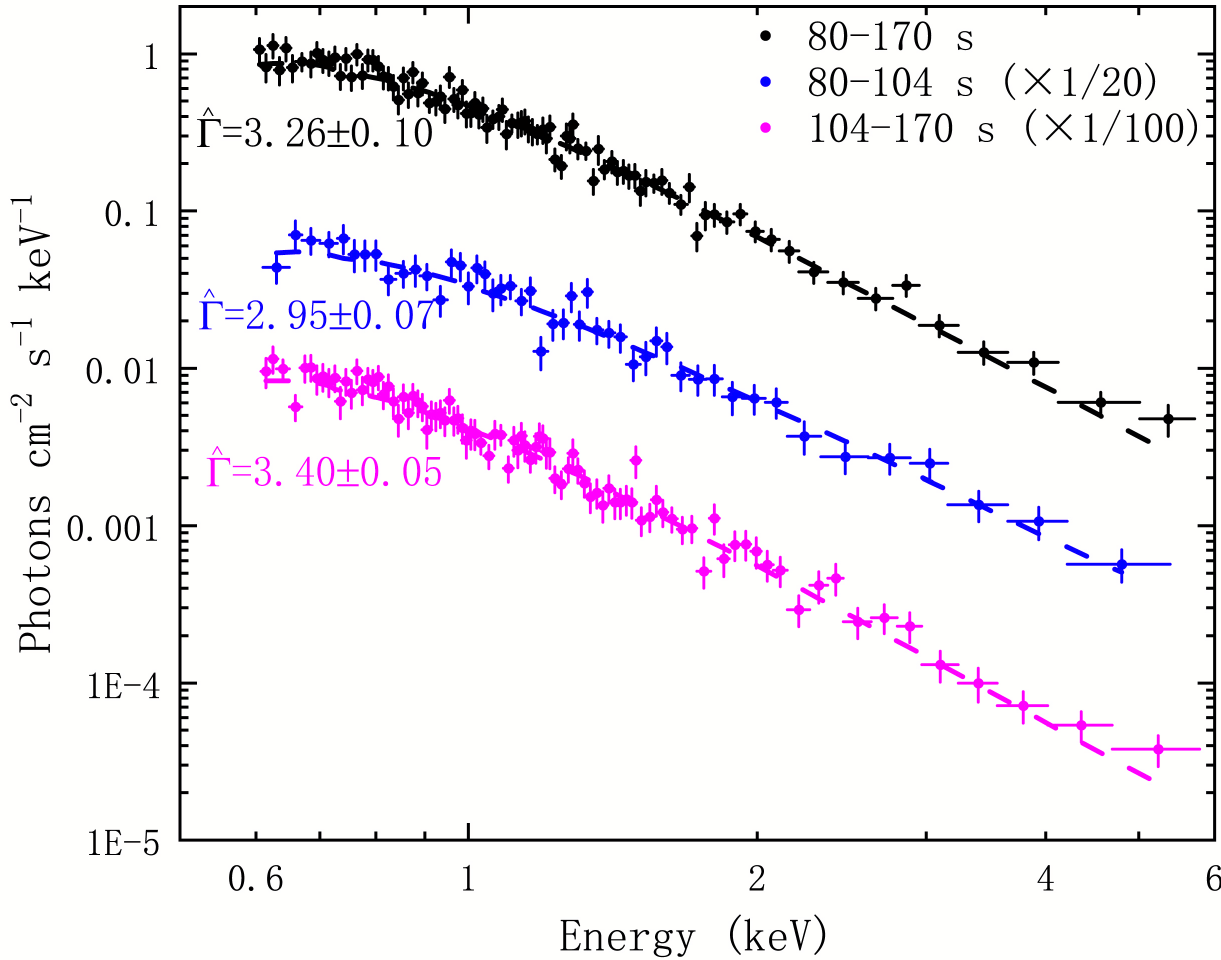}
\caption{Spectra with the PL model for the Pulse II (Flare I) segment of GRB 170519A.}
\label{fig:Flare1-spectra}
\end{figure}

\begin{figure}[htb!]
\centering
\includegraphics[angle=0,scale=0.35]{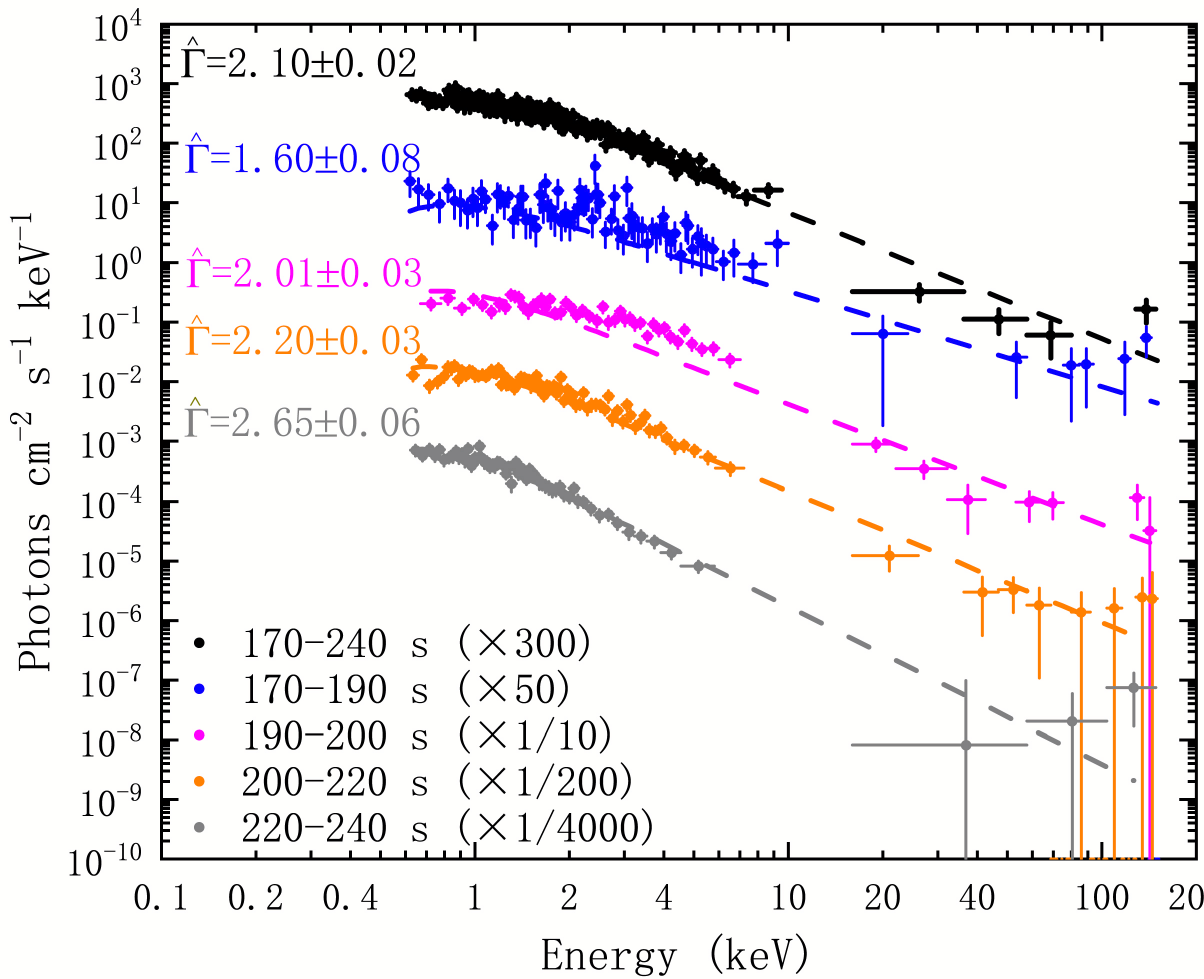}
\includegraphics[angle=0,scale=0.35]{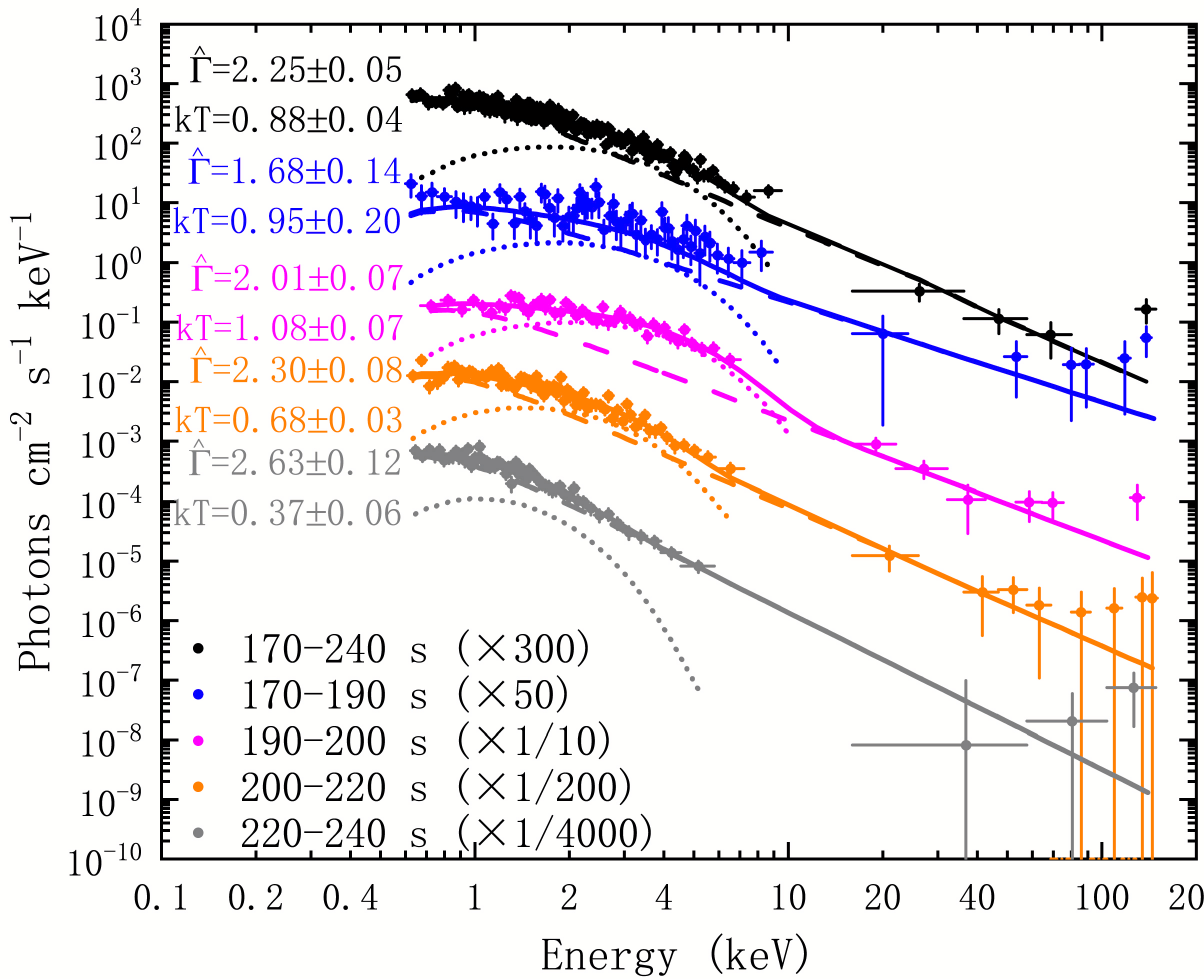}
\caption{Comparison of spectra with the PL (upper panel) and the BB+PL (lower panel) models during the Pulse III (Flare II) segment of GRB 170519A.}
\label{fig:Flare2-spectra}
\end{figure}

We fit the spectra using the PL model \citep[e.g.,][]{keith1999, peng2014}
\begin{equation}
\label{Spec-PL}
N_{\rm PL}=A_{\rm PL}E^{-\hat{\Gamma}},
\end{equation}
where $E$ is the observed energy, $A_{\rm PL}$ and $\hat{\Gamma}$ are the normalization and photon index of the PL spectrum, respectively. The fitting results of $\hat{\Gamma}$ with 1$\sigma$ uncertainties and the reduced (per degree of freedom, dof) chi-squared statistics ($\chi^2/{\rm dof}$) are listed in Table \ref{tab:SpectralFitting}. The analysis results of the three pulses and multiwavelength afterglow are as follows:
\begin{enumerate}
\item \textit{Pulse I:} We divide the pulse into three slices. As shown in Figure \ref{fig:BAT-spectra}, spectra of the first pulse are well fitted by the PL model, and the values of $\hat{\Gamma}$ exhibit a hard-to-soft evolution ($\hat{\Gamma}$ evolved from 1.29 to 2.13 over time). 

\item \textit{Pulse II (Flare I):} We divide the pulse into two slices. As shown in Figure \ref{fig:Flare1-spectra}, spectra of Pulse II are well fitted by the PL model. The values of $\hat{\Gamma}$ are 2.95 and 3.40 during $T_0$ + [80.0, 104.0] s and $T_0$ + [104.0, 170.0] s, respectively; the spectrum seems to become softer as time increases, and softer than spectra of the BAT pulse.

\item \textit{Pulse III (Flare II):} \textbf{The pulse is also jointly detected by BAT and XRT, and it can be seen as a giant X-ray flare with the soft $\gamma$-ray counterpart.} We divide the flare into four slices: $T_0$ + [170.0, 190.0] s, $T_0$ + [190.0, 200.0] s, $T_0$ + [200.0, 220.0] s, and $T_0$ + [220.0, 240.0] s. As shown in the top panel of Figure \ref{fig:Flare2-spectra}, we first use the PL model to fit their spectra. The data of the time-resolved spectra in the energy range of several keV are significantly higher than the fitted line, especially at $T_0$ + [190.0, 200.0] ($\chi^2$/dof = 271.19/103) and $T_0$ + [200.0, 220.0] ($\chi^2$/dof = 228.59/139). Therefore, an additional spectral component (e.g., BB component) should be added in the fit, as described in detail in Section \ref{sec:flare2}). 

\item \textit{Multiwavelength afterglow:} We divide the afterglow observed during $T_0$ + [$3.70\times 10^2$, $1.26\times 10^{5}$] s into three slices: $T_0$ + [$3.70 \times 10^2$, $8.00\times 10^2$] s (Slice 1), $T_0$ + [$4.15\times 10^3$, $6.48\times 10^3$] s (Slice 2), and $T_0$ + [$9.90 \times 10^3$, $3.60 \times 10^4$] s (Slice 3). Slice 1 characterizes the interval before the peak of the onset bump, while Slices 2 and 3 characterize the interval post bump (the range between Slice 2 and 3 is excluded since there is not enough data for analysis). As shown in Figure \ref{LC-BAT-XRT-OPT}, the joint optical and X-ray spectra are well fitted by the PL model. There is no obvious spectral evolution observed during the three slices. The spectral indices $\beta=\hat{\Gamma}-1$ of Slices 2 and 3 \textbf{are 0.86 and 0.92}, respectively. \textbf{We use their average value $\beta \approx 0.89$ in the following calculation.} 
For the fireball external-shock model \citep[e.g.,][]{sari1998, zhang2006, gao2013, wang2015} \textbf{with ISM medium, if $\nu_{\rm m}\le\nu_{\rm c}\le\nu_{\rm O}\le\nu_{\rm X}$ (where $\nu_{m}$ and $\nu_{c}$ respectively represent the minimum frequency and cooling frequency for synchrotron radiation), the electron index $p = 2\beta \approx 1.78 < 2$ and the expected temporal index $\alpha=(3\beta+5)/8=0.96$, which is consistent with the temporal indices $\alpha_{\rm X,1}=0.95\pm0.05$ and $\alpha_{\rm O,2}=0.92\pm0.01$. As a result, observations in Slices 2 and 3 are consistent with the predictions in the normal decay phase in the fireball external shock model.} 

\item \textit{Late $R$-band afterglow:} \textbf{In post-jet-break phase of the external-shock model with ISM medium, if $p=1.78$ and $\nu_{\rm m}\le\nu_{\rm c}\le\nu_{\rm O}\le\nu_{\rm X}$, a temporal decline $(3p+22)/16=1.71$ is predicted. As shown in Figure \ref{LC-BAT-XRT-OPT}, there are 4 $R$-band points after $T_0 + 1.26\times 10^5$ s decrease following $\propto t^{-1.71}$. Which indicates the decay of this late $R$-band light curve may originate from the radiation during the post-jet-break phase in the external shock model.}





\end{enumerate}

\tabletypesize{\scriptsize}
\begin{deluxetable*}{cccccccccc}
\centering 
\tablecaption{Spectral fitting with PL and BB+PL models of GRB 170519A\label{tab:SpectralFitting}.}
\tablewidth{1pt}
\tabcolsep=3pt
\tablehead{\multirow{3}{*}{Time interval (s)} & \multicolumn{3}{c}{PL}&\multicolumn{4}{c}{BB+PL}&\multirow{3}{*}{$\Delta$BIC}&\multirow{3}{*}{Preferred model}\\
 \cmidrule(r){2-4} \cmidrule(r){5-8} 
 & \colhead{$\hat{\Gamma}$} & \colhead{$\chi^2/$dof} & \colhead{BIC$_{\rm PL}$} & \colhead{$kT$ (keV)} & \colhead{$\hat{\Gamma}$} & \colhead{$\chi^2/$dof} & \colhead{BIC$_{\rm BB+PL}$} & & }
\startdata
\multicolumn{10}{c}{Pulse I} \\
$-50-50$&$ 1.87\pm0.12$& $55.51/56$ & $... $&...&...& ...&...&...&...\\
$-50--20$&$1.29\pm0.36$& $ 46.30/56$&$... $&...&...& ...&...&...&...\\
$-20-10$&$1.86\pm0.10$& $48.32/56 $&$... $&...&...& ...&...&...&...\\
$10-50$&$2.13\pm0.31$& $48.75/56 $&$... $&...&...& ...&...&...&...\\
\multicolumn{10}{c}{Pulse II (Flare I)}\\
$80-170$ & $3.26\pm0.10$ & $212.4/236$ & ... & ... & ... & ... & ... & ... & ...\\
$80-104$ &$2.95\pm0.07 $ & $47.31/47$ & ... & ... & ... & ... & ... & ... & ...\\
$104-170$&$3.40\pm0.05$ & $89.68/93$ & ... & ... & ... & ... & ... & ... & ...\\
\multicolumn{10}{c}{Pulse III (Flare II)}\\
$170-240$& $2.10\pm0.02$ & $381.73/234$ & $386.48$ &$0.88\pm0.04$ & $ 2.25\pm0.05$ &  $228.57/232$ & $238.06$&$148.41$&BB+PL (very strong) \\
$170-190$ & $1.60\pm0.08$& $204.46/305$ &$209.43$ & $0.96\pm0.20$ & $1.68\pm0.14$ & $ 109.94/303$ & $209.89$&$-0.45$&PL \\
$190-200$ &$2.01\pm0.03$ & $273.47/101$&$277.50$ &$1.08\pm0.07$ &$2.01\pm0.07$ & $77.43/99$& $85.38$&$192.01$&BB+PL(very strong)\\
$200-220$ &$2.20\pm0.03$& $225.39/135$ &$229.66$ &$0.68\pm0.03$ &$2.30\pm0.08$ & $143.98/133$ &$152.53$&$77.14$& BB+PL (very strong)\\
$220-240$&$2.65\pm0.06$ & $88.90/116$ &$93.04$ &$0.37\pm0.06$ &$ 2.63\pm0.12$ & $76.82/114$&$85.11$&$7.94$&{BB+PL (strong)}\\
\multicolumn{10}{c}{The afterglow observed after $T_0$ + 370 s}\\
slice 1: $370-800$&$ 1.86\pm0.02$& $43.94/16$&...&...&...& ...&...&...&...\\
slice 2: $4153-6483$&$1.86\pm0.01$& $ 394.33/38$&...&...&... &...&...&...&...\\
slice 3: $9895-35978$&$1.92\pm0.01$ & $280.00/38$ &...&...&...& ...&...&...& ...\\
\enddata

\end{deluxetable*}

\begin{figure}[htb!]
\centering
\includegraphics[angle=0,scale=0.35]{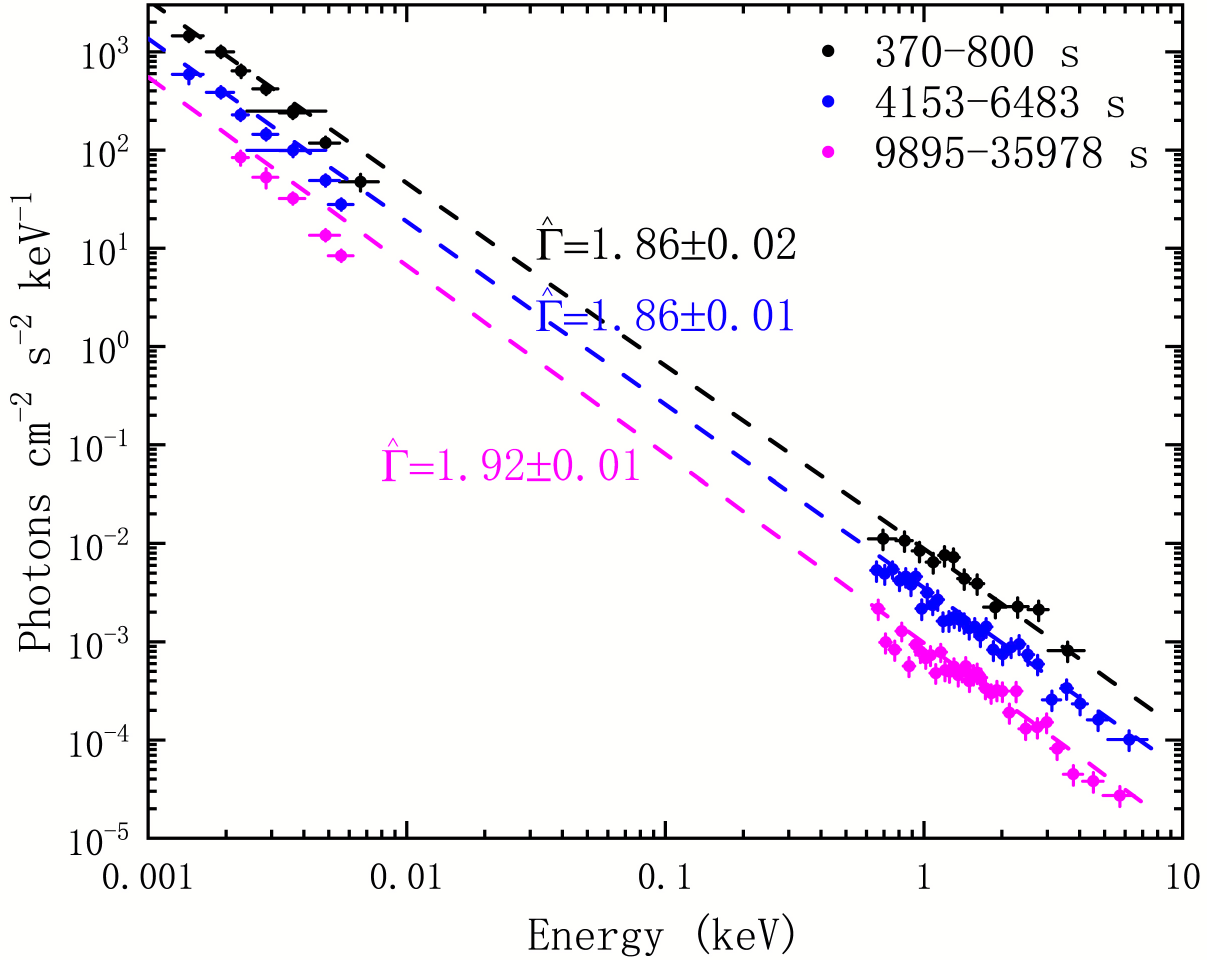}
\caption{Spectra with the PL model for the afterglow of GRB 170519A, for Slices I, II, and III.}
\label{fig:M-aftergow}
\end{figure}

\subsection{Thermal Radiation in Flare II}
\label{sec:flare2}
For both the time-integrated and time-resolved spectra of the Pulse III (Flare II), deviations are observed between the data and the fit at several keV, indicating an additional spectral component may be superimposed on the PL component. In the framework of the fireball model, the photospheric emission with BB spectrum could be involved in the matter-dominant outflow \citep[e.g.,][]{merzaros2000, Falcone2007, Page2011, peng2014}. In order to identify the potential BB component, we consider the BB function (Planck function) + PL as an alternative model \citep[e.g.,][]{keith1999, peng2014},
\begin{equation}
\label{Spec-PL+BB}
N_{\rm BB+PL}=A_{\rm BB}\frac{8.0525\,E^2dE}{(kT)^4(e^{E/kT}-1)}+A_{\rm PL}E^{-\hat{\Gamma}},
\end{equation}
where $A_{\rm BB}$ and $kT$ are the normalization and temperature of the BB model, respectively. To evaluate the goodness of fit of the PL and BB+PL models, we calculated the Bayesian information criterion (BIC) \citep[e.g.,][]{kass1995, trotta2008} for each of them. Through the definition of $\Delta{\rm BIC} = {\rm BIC}_{\rm PL}-{\rm BIC}_{\rm BB+PL}$ (where ${\rm BIC}_{\rm model}$ represents the BIC value of the model), the preferred model is determined based on the sign of $\Delta$BIC: a positive value of $\Delta$BIC indicates that the model BB+PL is preferred. The significance level of the selected model is determined based on the value of $\Delta {\rm BIC}$ \citep[e.g.,][]{trotta2008, Wilson2013EmbracingBF, hou2018, qin2021, song2024}: (1) if $0<\Delta {\rm BIC}<2$, the preference for model BB+PL is not worth more than a bare mention; (2) if $2<\Delta {\rm BIC}<6$, model BB+PL is preferred; (3) if $6<\Delta {\rm BIC}<10$, model BB+PL is strongly supported; (4) if $\Delta {\rm BIC} > 10$, the preference for model BB+PL is very strong. The spectral fitting results, ${\rm BIC}_{\rm PL}$, ${\rm BIC}_{\rm BB+PL}$, and $\Delta$BIC are listed in Table \ref{tab:SpectralFitting}. The spectra with the PL and BB+PL models are shown in Figure \ref{fig:Flare2-spectra}. 
One can see that the time-integrated spectrum and the time-resolved spectra during the flare favor the BB+PL model; the blackbody temperature $kT$ decays from $1.08\pm0.07$ to $0.37\pm0.06$ keV. During the slice $T_0$ + [170.0, 190.0] s, the X-ray flare was not triggered, and the $\Delta{\rm BIC}$ value does not provide clear support for the BB+PL model. Therefore, the thermal radiation may be generated concurrently with the X-ray flare\footnote{\textbf{Noted that \cite{valan2021} presented a search for significant BB components in 116 GRBs (included GRB 170519A), using a high confidence level $> 3 \sigma$ and $\Delta 
\chi^2\ge 2$ which should be simultaneously obtained in three consecutive slices, they did not find significant BB component in GRB 170519A. Here, we search the potential BB using BIC method, which have been widely used in previous BB component investigations \citep[e.g.,][]{li2019, lirq2022, chang2023, chang2024}) }}. 



\startlongtable
\begin{deluxetable*}{cccccccc}
\centering
\setlength{\tabcolsep}{3pt}
\tablecaption{Photospheric parameters for GRB 170519A} \label{tab:bb_parameters}
\tablehead{\colhead{Slice} & $f_{\rm tot}$ & $f_{\rm BB}$ & $E_{\rm iso}$ &\colhead{$L_{\rm BB}$} & \colhead{$R_{\rm ph}$} & \colhead{$\Gamma_{\rm BB}$} & $R_0$ \\
 (s) & ($\times 10^{-8}$ erg s$^{-1}$ cm$^{-2}$) & ($\times 10^{-9}$ erg s$^{-1}$ cm$^{-2}$) & ($\times 10^{51}$ erg) & ($\times 10^{49}$ erg s$^{-1}$) & ($\times 10^{11}$ cm) & & ($\times 10^{7}$ cm) }
\startdata
$190-200$ & $3.86 \pm 0.32$ & $20.40 \pm 1.69$ & $1.23\pm 0.10$ & $3.57 \pm 0.30$ & $2.32 \pm 1.04$ & $67.71 \pm 9.90$ & $6.66 \pm 0.95$\\
$200-220$ & $1.94 \pm 0.29$ & $6.50 \pm 0.95$ & $1.24\pm 0.18$ & $1.14 \pm 0.17$ & $2.57 \pm 1.35$ & $52.09 \pm 8.78$ & $4.84 \pm 1.22$\\
$220-240$ & $0.79 \pm 0.36$ & $1.37 \pm 0.63$ & $0.50\pm 0.23$ & $0.24 \pm 0.11$ & $1.44 \pm 1.17$ & $46.70 \pm 10.49$ & $0.73 \pm 0.58$\\
\enddata
\end{deluxetable*}

\begin{figure}[htb!]
\hspace{-25pt}
\includegraphics[angle=0,scale=0.46]{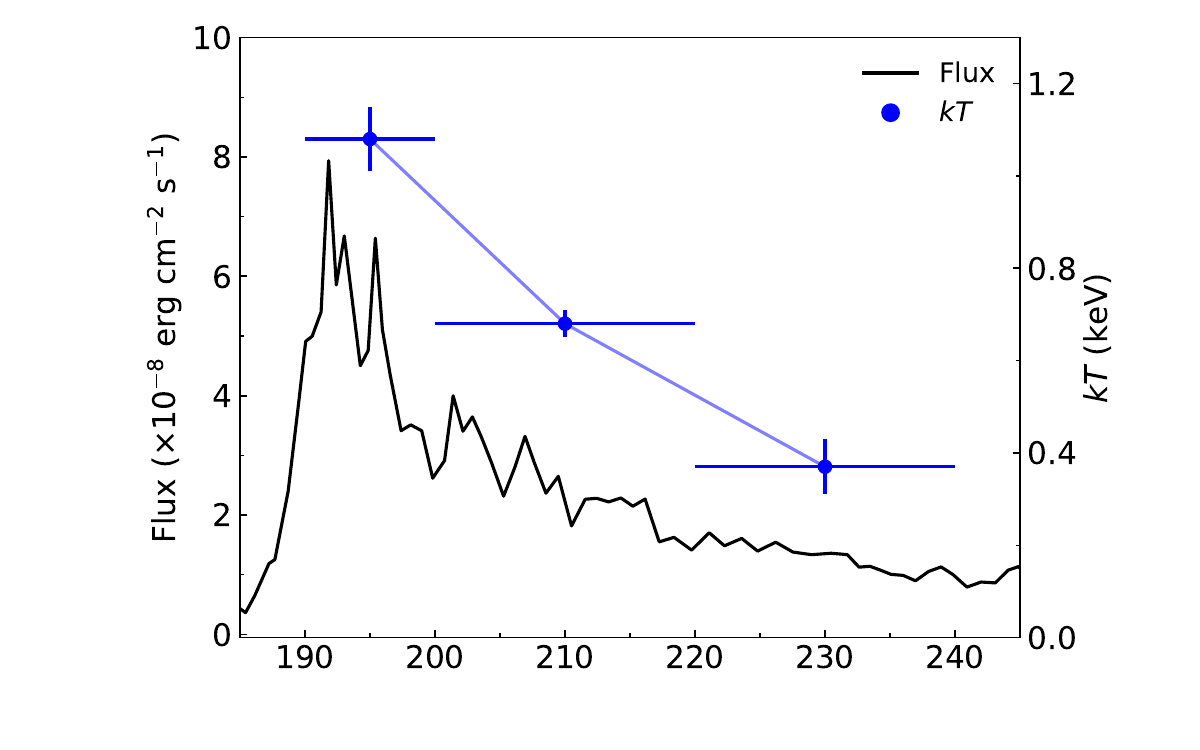} 
\includegraphics[angle=0,scale=0.5]{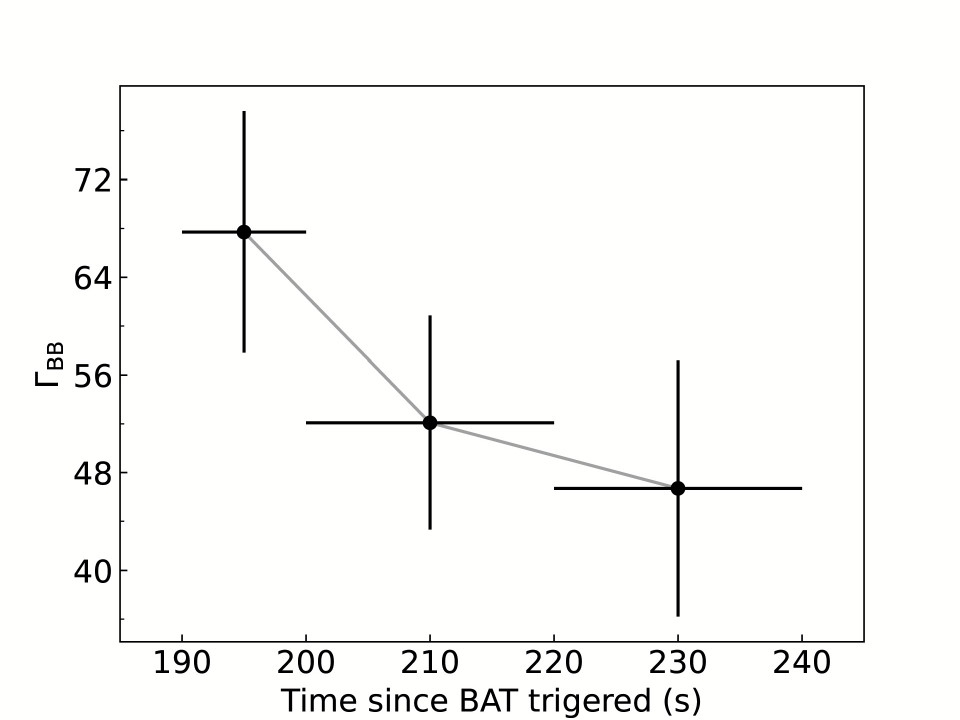} 
\includegraphics[angle=0,scale=0.5]{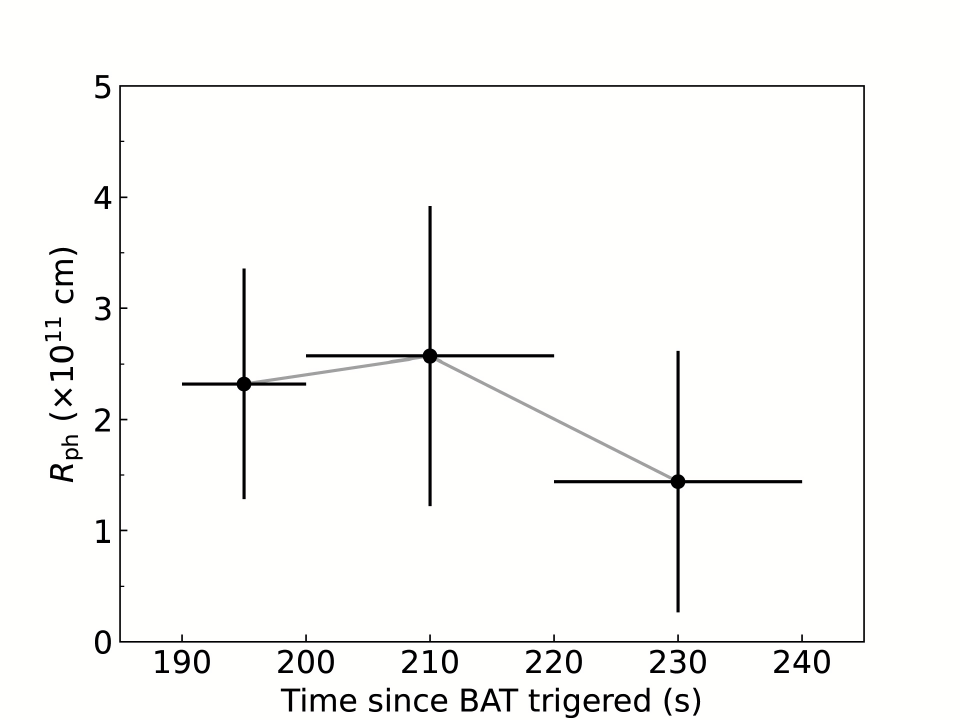} 
\caption{Evolution of $kT$ (top panel), $\Gamma_{\rm BB}$ (middle panel) and $R_{\rm ph}$ (bottom panel).}
\label{fig:param_evolution}
\end{figure}

\textbf{In the framework of the blackbody model in a fireball, the photosphere radius $R_{\rm ph}$, the Lorentz factor $\Gamma_{\rm BB}$ and the central engine radius $R_0$ 
can be estimated from the observation (total flux $f_{\rm tot}$, BB flux $f_{\rm BB}$ and $kT$
) \citep[e.g.,][]{meszaros2002, peer2007, fan2012, gao2015, zhangbook18}. The values of $f_{\rm tot}$, $f_{\rm BB}$, $E_{\rm iso}$ and $L_{\rm BB}$ could be calculated from the different time-resolved spectra spectrum, and the results of GRB 170915A are listed in Table \ref{tab:bb_parameters} and. Assuming $R_{\rm ph}>R_{\rm s}$ ($R_{\rm s}$ represents the saturated radius)\footnote{\textbf{We also tried to consider the situation of $R_{\rm s}>R_{\rm ph}$, the result of saturated Lorentz factor $\eta<\eta^*$ contradicts the assumption}}, we obtained the values of $\Gamma_{\rm BB}$, $R_{\rm ph}$ and $R_0$, and also shown in Table \ref{tab:bb_parameters} Figure \ref{fig:param_evolution}. In our results, $R_0\sim 10^7$ cm is consistent with typical size of central engine \citep[e.g.,][]{paczynski1986, preece2002, ryde2010, peng2014, zhangbook18}. During decaying of flux, $kT$ and $\Gamma_{\rm BB}$ also decrease with time, and the values of $R_{\rm ph}$ are around $2\times10^{11}$ cm. }


\begin{figure}[htb!]
\vspace{+0.1cm} 
\centering
\includegraphics[angle=0,scale=0.5]{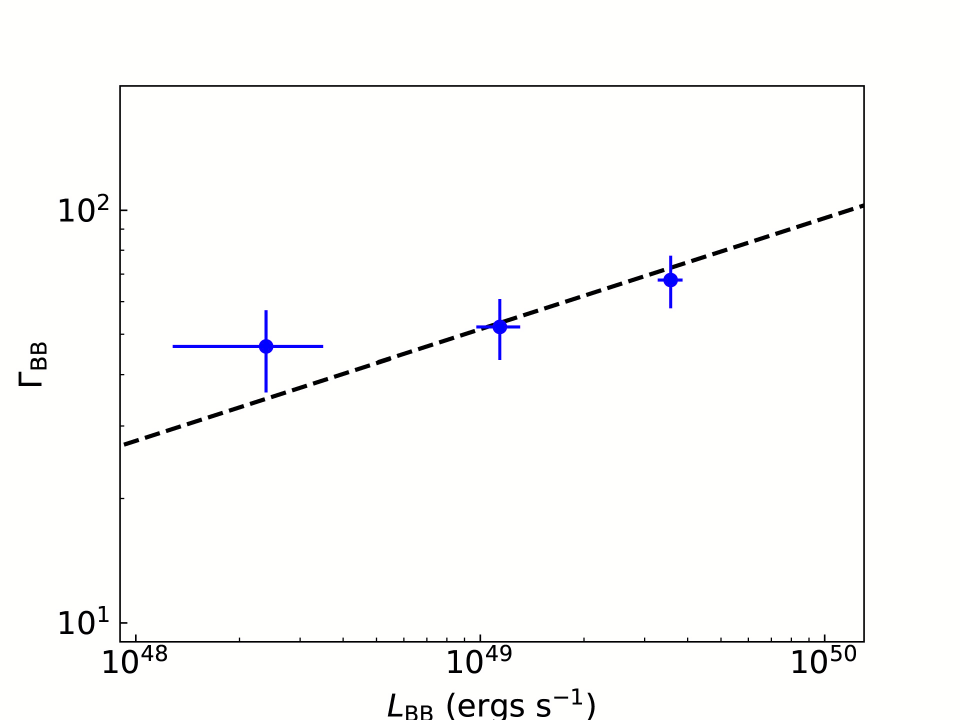}
\includegraphics[angle=0,scale=0.5]{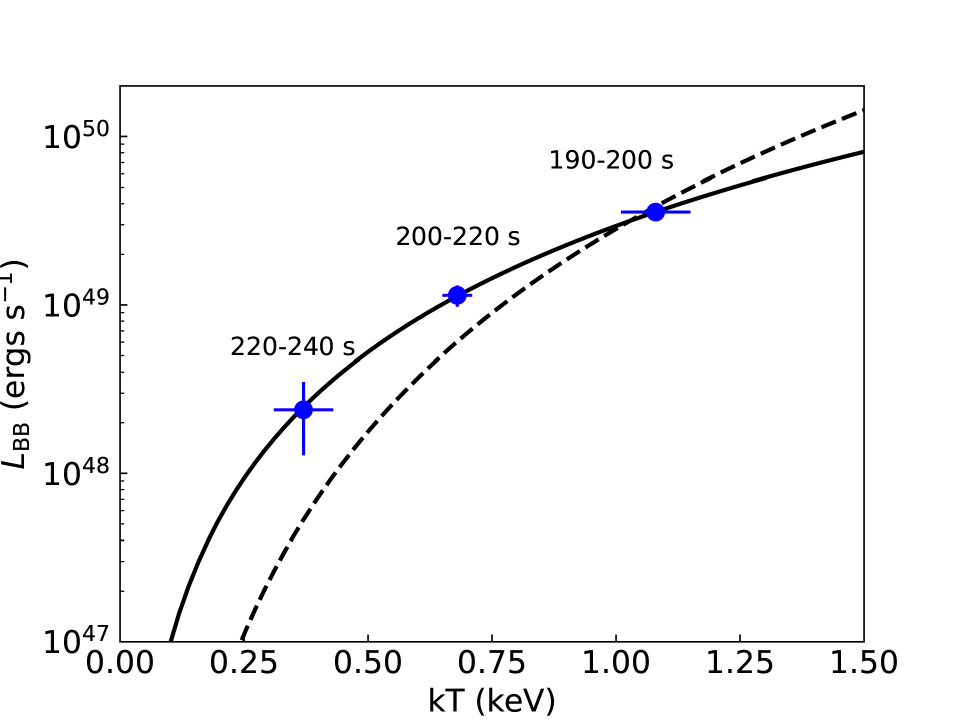} \\
\caption{\textbf{Relations between photospheric parameters during Pulse III (Flare II), where the slices with $\Delta$BIC $>2$ are included. Top panel shows the $L_{\rm BB}$--$\Gamma_{\rm BB}$ plane for Pulse III (Flare II), and the dashed line represents the relation $\Gamma_{\rm BB}\propto L_{\rm BB}^{0.27}$. Bottom panel: Relation between $kT$ and $L_{\rm BB}$, where the solid line represents the fitting result $L_{\rm BB}\sim kT^{2.54\pm0.17}$, and the dashed line represents $L_{\rm BB}\sim kT^{4}$.
}}
\label{fig:relation}
\end{figure}

\textbf{The relations of $L_{\rm BB}$--$\Gamma_{\rm BB}$ and $kT$--$L_{\rm BB}$ are shown in Figure \ref{fig:relation}. \cite{fan2012} suggested that blackbody emissions from a baryonic photosphere can produce a relation of $\Gamma_{\rm BB}\propto L_{\rm BB}^{0.27}$ (where $\Gamma_0\approx\Gamma_{\rm BB}\approx\eta$ was assumed). As shown in the top panel of Figure \ref{fig:relation}, our results well follow the $\Gamma_{\rm BB}\propto L_{\rm BB}^{0.27}$ relation, which is consistent with previous statistical studies \citep[e.g.,][]{fan2012, peng2014, valan2021}. We also fitted the $kT$--$L_{\rm BB}$ relation using a PL function, and found that}
\begin{equation}
\label{kT-LBB}
\bf
L_{\rm BB}=(2.95\pm0.02)\times10^{49}\,\left(\frac{kT}{1\ \rm keV}\right)^{2.49\pm0.03} {\rm erg/s}.
\end{equation}
\textbf{The observed $kT$--$L_{\rm BB}$ relation is shallower than $kT^4$, which may be} due to the mild evolution of $\bf \Gamma_{\rm BB}$ and \textbf{consistent with the previous results \citep[e.g.,][]{hou2018,valan2021}.} On the other hand, the $\hat{\Gamma}$ values in the flare, which represent the non-thermal parts of spectra, also exhibit hard-to-soft evolution (evolving from 1.60 to 2.65 over time). The spectral evolution of both flares and the prompt gamma-ray emission may suggest a common origin \citep[e.g.,][]{barthelmy2005, zhang2006}.

\subsection{Modeling with External Shock Model}
\label{sec:modeling}
As shown in Section \ref{sec:spectral}, the normal decay of GRB 170519A supports the prediction of multiwavelength synchrotron radiation produced by the slow cooling of the external shock wave, in the regime $\nu_m<\nu_c<\nu_O<\nu_X$ regime. \textbf{The forward shock emission and the reverse shock emission for GRBs standard afterglow model has been derived in detail by \citet{sari1998} and \citet{zhang2005}, respectively. In our work, we calculated numerically of the dynamical evolution of the shell using the hydrodynamical equations from \citet{HuangYongFen1999} and \citet{gao2013}, and calculated the contribution of forward shock emission with \citet {sari1998}. }
In this model, the relativistic shell with isotropic kinetic energy $E_{\rm K,iso}$ and initial Lorentz factor $\Gamma_0$ colliding with the external medium, a relativistic shock is generated and propagates through a uniform cold medium with particle density $n$. In the shocked regime, electrons may be accelerated, and they generate a multiwavelength afterglow through synchrotron radiation. Assuming the accelerated electrons follow a power-law spectrum with index $p$, the fraction of shock energy to electron energy is $\epsilon_e$, the fraction of shock energy to magnetic field energy is $\epsilon_B$. \textbf{We set $p$ = $2\beta$=1.78, located in $\nu_c<\nu_O<\nu_X$ regime and ISM medium.} 

\begin{figure*}[htb!]
\centering
\includegraphics[angle=0,scale=0.5]{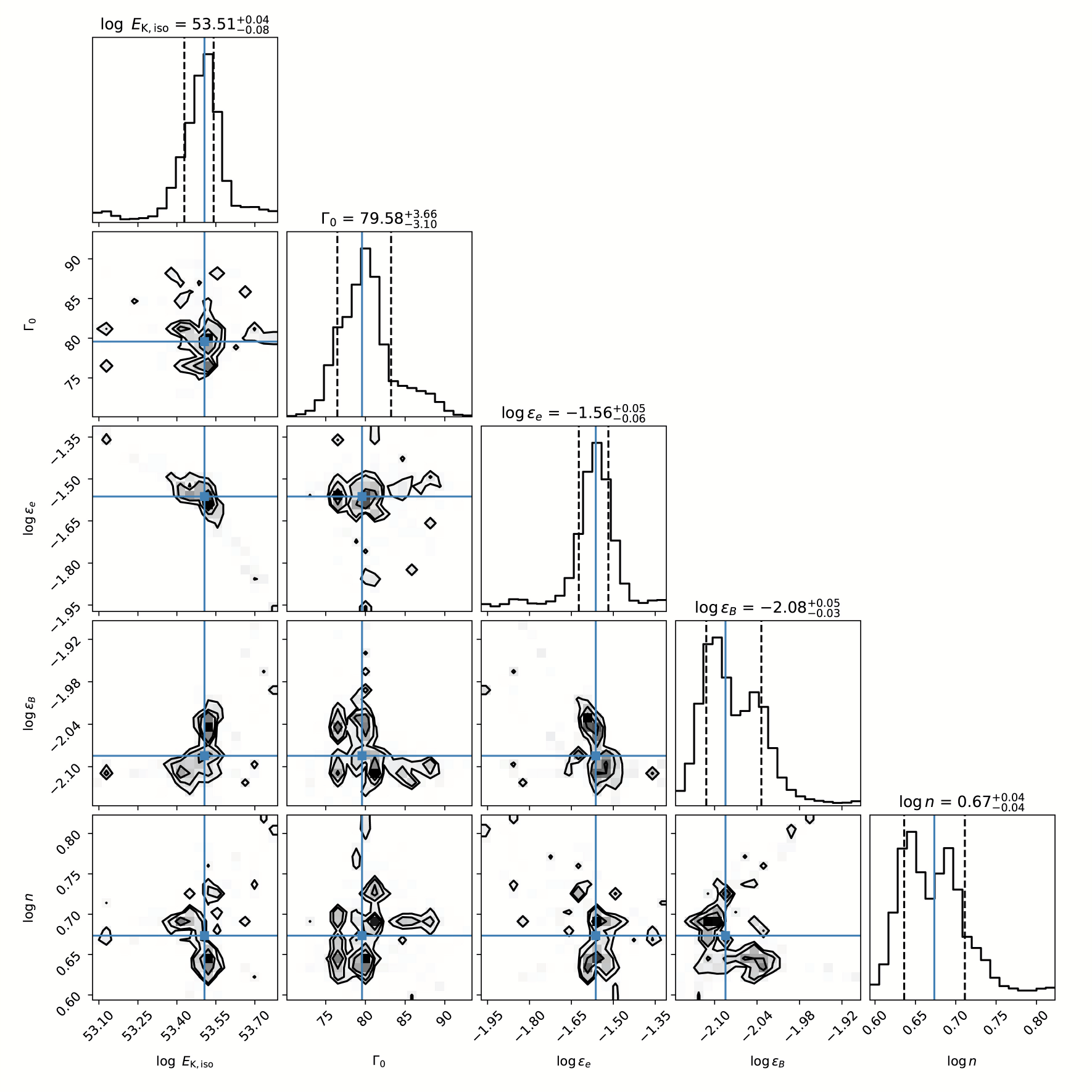}
\caption{Corner plots of our MCMC parameters for external forward model in ISM medium. The uncertainties are computed within 1$\sigma$ confidence ranges, whose boundaries are shown as dashed lines, the best-fit parameters are shown as solid lines.}
\label{fig:mcmc}
\end{figure*}

\begin{figure}[htb!]
\centering
\includegraphics[angle=0,scale=0.5]{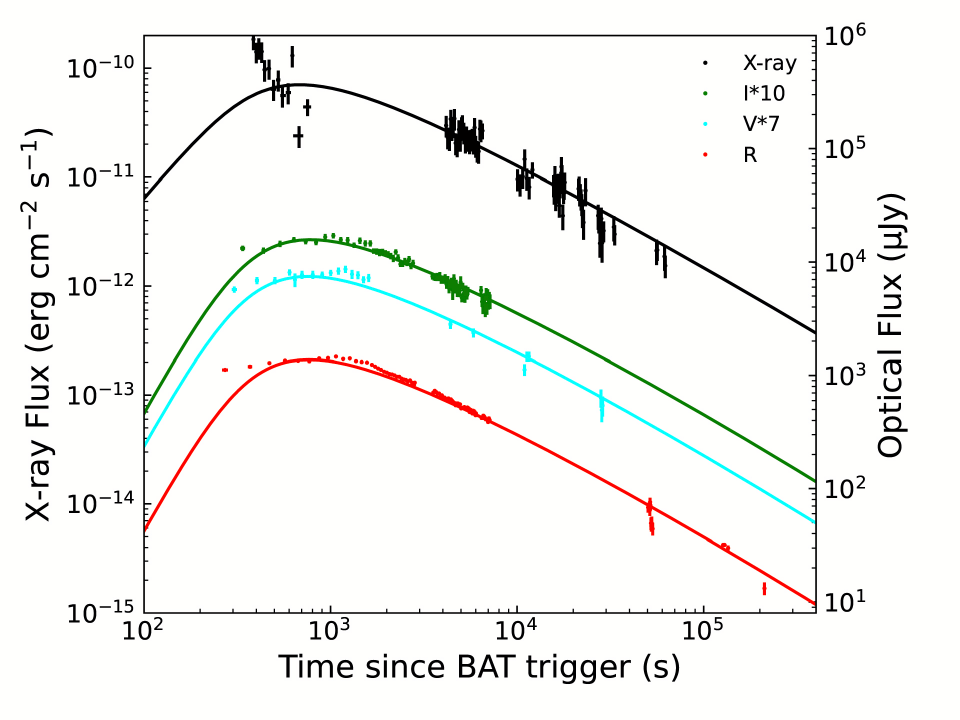}
\caption{Light curve of the GRB 170519A afterglow and the fitting with the standard external-shock model.}
\label{fig:FS}
\end{figure}

\textbf{To search for the best-fit parameter set we obtain the parameters using the Markov Chain Monte Carlo (MCMC) method, which is performed using the emcee Python package \citep{foreman2013}, with 25 walkers and 500 steps. The free parameters and fitting ranges for this model are $E_{\rm K,iso} \in [10^{52}, 10^{56}]$ erg, $\log \Gamma_{0} \in [1.5, 2.5]$, $\log \epsilon_{\rm e} \in [-3,-0.5]$, $\log \epsilon_{\rm B} \in [-5,-2]$, $n \in [10^{-2}, 10^{2}]$ cm$^{-3}$. When $T > T_0 + 10^3$ s, the best-fitting parameters of the standard external forward shock model are $n = 4.71^{+0.43}_{-0.39}$ cm$^{-3}$, $E_{\rm K, iso}=3.21^{+0.28}_{-0.53}\times10^{53}$ erg, $\Gamma_0=79.58^{+3.66}_{-3.10}$, $\epsilon_e=2.74^{+0.30}_{-0.36}\times10^{-2}$ and $\epsilon_B=8.23^{+1.02}_{-0.50}\times10^{-3}$, with $\chi^2$/dof=4.60. The corner plots to of our MCMC parameter estimates are shown in Figure \ref{fig:mcmc}, the light curve and fitting of this model is shown in Figure \ref{fig:FS}. We can see that they can well constrain the observational data when $T > T_0 + 10^3$ s. }

\begin{figure}[htb!]
\centering
\includegraphics[angle=0,scale=0.5]{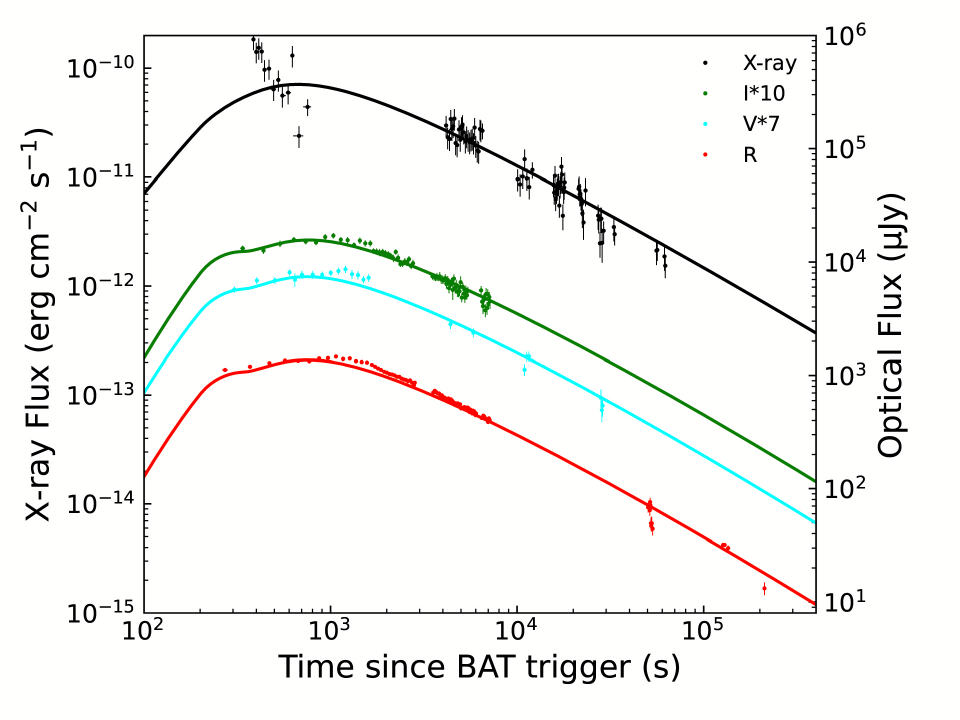}
\caption{\textbf{Light curve of the GRB 170519A afterglow and the fitting external shock model with time-dependent $\epsilon_B$.}}
\label{fig:FS-eB-e}
\end{figure}

\textbf{Usually, the microphysical parameters, e.g., $\epsilon_e$ and $\epsilon_B$, in the standard model are typically assumed to be not varying, and they are consistent with the observations of late-time afterglows \citep[e.g.,][]{Yost2003,wang2015, wang2018, eftekhari2020}. However, the mechanism of energy transfer from protons to electrons and magnetic fields in the relativistic shocks is complicated. The time-dependent $\epsilon_B$ ($\epsilon_B \propto t^{k}$) in standard afterglow model have been proposed to solve some difficulties encountered with observations, such as early phase afterglow, chromatic breaks \citep[e.g.,][]{Ioka2006,panaitescu2006,Kong2010,vanderhost2014,Huang2018, Fraija2024}. The early time emission is even more complicated. We therefore assume that $\epsilon_B$ evolves with time, while the other parameters are the same as above. As shown in Figure \ref{fig:FS-eB-e}, the flattening of early optical afterglow can be well fitted by considering the evolution of $\epsilon_B \propto t^{-3.19}$, with $\chi^2$/dof=2.35. The value of $\epsilon_B$ in the early emission epoch ($T < T_0 + 10^3$ s) evolves from $2.74\times 10^{-2}$ to $8.23 \times 10^{-3}$.}

\section{Conclusions and Discussion}  
\label{sec:conclusions}
GRB 170519A was detected by \emph{Swift}/BAT. Its broadband afterglow was detected by \emph{Swift}/XRT, \emph{Swift}/UVOT, and ground-based optical telescopes. The well-sampled optical light curves were acquired from $\rm{T_0} + 272$ s to 3.4 days after the \emph{Swift}/BAT trigger. We report Lick/KAIT observations of GRB 170519A, and investigate the physical origins of the multiwavelength afterglow.

Through the temporal analysis and joint spectral analysis, the properties of \textbf{GRB 170519A} are summarized as follows.
\begin{enumerate}
\item There are three pulses during $T_0$ + [$-50.0$, 240.0] s, including a pulse detected only by BAT, an X-ray flare (Pulse II/Flare I), and a bright flare (Pulse III/Flare II) detected by both BAT and XRT. It is found that the spectral evolution of all three pulses is hard-to-soft by fitting with the PL model.
\item There are deviations between the data and the PL fittings of the spectra of Pulse III (Flare II). We use the model BB+PL to fit its time-integrated spectrum and time-resolved spectra, and find that the BB+PL model is favored at $T_0$ + [190.0, 240.0].

\item \textbf{Using the measurements from spectra of Pulse III (Flare II), we estimate the initial radius of fireball $R_0\sim 10^7$ cm, the photospheric radius $R_{\rm ph}\sim 10^{11}$ cm. The temperature $kT$ decreasess with time \textbf{from 1.08 to 0.37 keV}, and Lorentz factor of blackbody $\Gamma_{\rm BB}$ decreases with time from $67.71$ to $46.70$.} 

\item \textbf{We find that $\Gamma_{\rm BB}$ and $L_{\rm BB}$ of GRB 170519A follows the relation $\Gamma_{\rm BB} \propto L_{\rm BB}^{0.27}$, which is expected from \cite{fan2012} and indicates the blackbody component of GRB 170519A is generated from the baryonic photosphere in fireball. We also find $\bf L_{\rm BB} \approx kT ^{2.49\pm 0.03}$, which is consistent with the evolutions of $\Gamma_{\rm BB}$.}

\item There is an onset bump well observed by KAIT in the early afterglow, rising with an index of $-0.43$ and peaking at $\sim T_0 + 1174.9$ s. It then exhibits the normal decay with an index $\alpha_{O,2}=0.92$ for the optical light curve and $\alpha_{X,1}=0.95$ for the X-ray light curve, respectively. In the normal decay phase, the photon index $\hat{\Gamma}$ is estimated as $\sim 1.9$ and no obvious evolution is found. The spectral and temporal indices\textbf{ of late afterglow} satisfy the closure relation prediction of synchrotron radiation from an external shock \citep[e.g.,][]{wang2015}.

\item We model the multiwavelength light curve of GRB 170519A using the external-shock model with time-dependent $\epsilon_B$. In the early afterglow \citep[e.g.,][]{Ioka2006}, the value of $\epsilon_B$ decays rapidly from $\bf 4.29\times10^{-2}$ to $\bf 8.23\times10^{-3}$,\textbf{ causing a flatter onset.} The best-fit parameters are $\bf n = 4.71$ cm$^{-3}$, $\bf E_{\rm k,iso}=3.21\times10^{53}$ erg, $\bf \Gamma_0=79.58$ and $\bf \epsilon_e=0.03$ for $\bf p=1.78$ \textbf{in $\nu_m<\nu_c<\nu$ regime}.
\end{enumerate}

\begin{acknowledgments}
This work makes use of data supplied by the UK \emph{Swift} Science Data Centre at the University of Leicester, and is supported by the National Natural Science Foundation of China (Grant No. 12373042, U1938201, 12133003), the Progamme of Bagui Scholars Programme (W.X.-G.), the Guangxi Science Foundation (grant 2018GXNSFGA281007).
A.V.F.’s research group at UC Berkeley acknowledges financial assistance from the Christopher R. Redlich Fund, Gary and Cynthia Bengier, Clark and Sharon Winslow, Alan Eustace (W.Z. is a Bengier-Winslow-Eustace Specialist in Astronomy), and numerous other donors.
KAIT and its ongoing operation were made possible by donations from Sun Microsystems, Inc., the Hewlett-Packard Company, AutoScope Corporation, Lick Observatory, the U.S. National Science Foundation, the University of California, the Sylvia \& Jim Katzman Foundation, and the TABASGO Foundation. Research at Lick Observatory is partially supported by a generous gift from Google.

\end{acknowledgments}

\bibliography{ref}{}
\bibliographystyle{aasjournal}

\end{document}